\documentclass[12pt]{article}
\usepackage{graphicx}
\begin{document}

\pagestyle{empty}

\noindent
{\bf Sparsely-Observed Pulsating Red Giants in the AAVSO Observing Program}

\bigskip

\noindent
{\bf John R. Percy\\Department of Astronomy and Astrophysics, and\\Dunlap Institute of Astronomy and Astrophysics\\University of Toronto\\Toronto ON\\Canada M5S 3H4}

\bigskip

{\bf Abstract}  This paper reports on time-series analysis of 156 pulsating
red giants (21 SRa, 52 SRb, 33 SR, 50 Lb) in the AAVSO observing program
for which there are no more than 150-250 observations in total.  Some
results were obtained for 68 of these stars: 17 SRa, 14 SRb, 20 SR, and 17 Lb.
These results generally include only an average period and amplitude.
Many, if not most of the stars are undoubtedly more complex; pulsating red
giants are known to have wandering periods, variable amplitudes, and often
multiple periods including ``long secondary periods" of unknown origin.
These results (or lack thereof) raise the question of how the AAVSO should best manage the observation of
these and other sparsely-observed pulsating red giants.

\medskip

\noindent
AAVSO keywords = AAVSO International Database; photometry, visual; pulsating variables; giants, red; period analysis; amplitude analysis

\medskip

\noindent
ADS keywords = stars; stars: late-type; techniques: photometric; methods: statistical; stars: variable; stars: oscillations

\medskip

\noindent
{\bf 1. Introduction}

\smallskip

Red giants are unstable to radial pulsation.  As they expand and cool,
the period and amplitude of pulsation increase.  Pulsating
red giants (PRGs) are classified as Mira if they have well-pronounced
periodicity and visual amplitude greater than 2.5, as SRa if they
have smaller amplitudes and persistent periodicity, as SRb if the
periodicity is poorly expressed, and Lb if the variability is irregular.
These classes are arbitrary; there is a spectrum of behavior from strictly
periodic to completely irregular, and of amplitudes from millimagnitudes
up to 10 magnitudes.

The AAVSO International Database contains observations of thousands
of PRGs.  Some are well-studied, especially the brighter Miras; see
Templeton {\it et al.} (2005) for a study of the periods and period
changes in 547 of them.  There are many, however, which have not been
studied, often because the number of observations is insufficient.  A few
years ago, my students and I undertook a study of some PRGs
for which there were only a few hundred observations: SRa/SRb/SR stars
(Percy and Tan 2013, Percy and Kojar 2013) and Lb stars (Percy and Long
2010, Percy and Terziev 2011).

The present paper describes a study of several dozen more SR and Lb
stars for which there were a total of 150-250 observations, and for
which analysis might be possible.  I thank Elizabeth Waagen, at AAVSO
HQ, for compiling lists of these sparsely-observed SRa, SRb, SR, and
Lb stars.  Although the primary purpose of this study was to determine
the basic variability parameters of as many of these stars as possible,
an equally-important purpose was, more generally, to determine whether sparsely-observed
PRGs can yield any meaningful results.

\medskip

\noindent
{\bf 2. Data and Analysis}

\smallskip

Observations were taken from the AAVSO International Database (Kafka
2018).  They ranged from all visual for some stars, to all Johnson V
(photoelectric or CCD) for others.  Periods were determined (or searched
for) using the Fourier routine in the VSTAR software package (Benn 2013).
Some of the stars had been studied with the All-Sky Automated Survey
(ASAS: Pojmanski 1997), and a period had been derived.  In many cases, the
ASAS light curve showed that the variability was complex, and occurred
on two or more periods or time scales.  This may be true for most of
our stars.

\medskip

\noindent
{\bf 3. Results}

\smallskip

Tables 1-4 list results for the stars classified as SRa, SRb, SR, and
Lb, respectively.  Columns list: the star; the period in the VSX catalog
(PVSX); the mean period and semi-amplitude obtained in the present study;
and notes about the star.  Visual amplitudes are denoted v; Johnson
V amplitudes as V.  The notes are as follows: 1: new period gives a
better phase curve than PVSX; 2: new period and PVSX give equally good
phase curves; 3: PVSX gives a poor phase curve; 4: neither new period
or PVSX gives a good phase curve; *: see note in section 3.1.  Figure 1
gives one example: of an SR star (EQ And) which shows a quite acceptable
phase curve.

Many of the Lb stars in Table 4 were observed primarily in Johnson V,
and produce acceptable results with only a few dozen observations.

The following are the number of stars analyzed, and the number and
percent which produced results, and which appear in Tables 1-4: SRa: 21, 17, 81\%; SRb: 52, 14,
27\%; SR: 33, 20, 61\%; Lb: 50, 17, 34\%.

\begin{figure}
\begin{center}
\includegraphics[height=7cm]{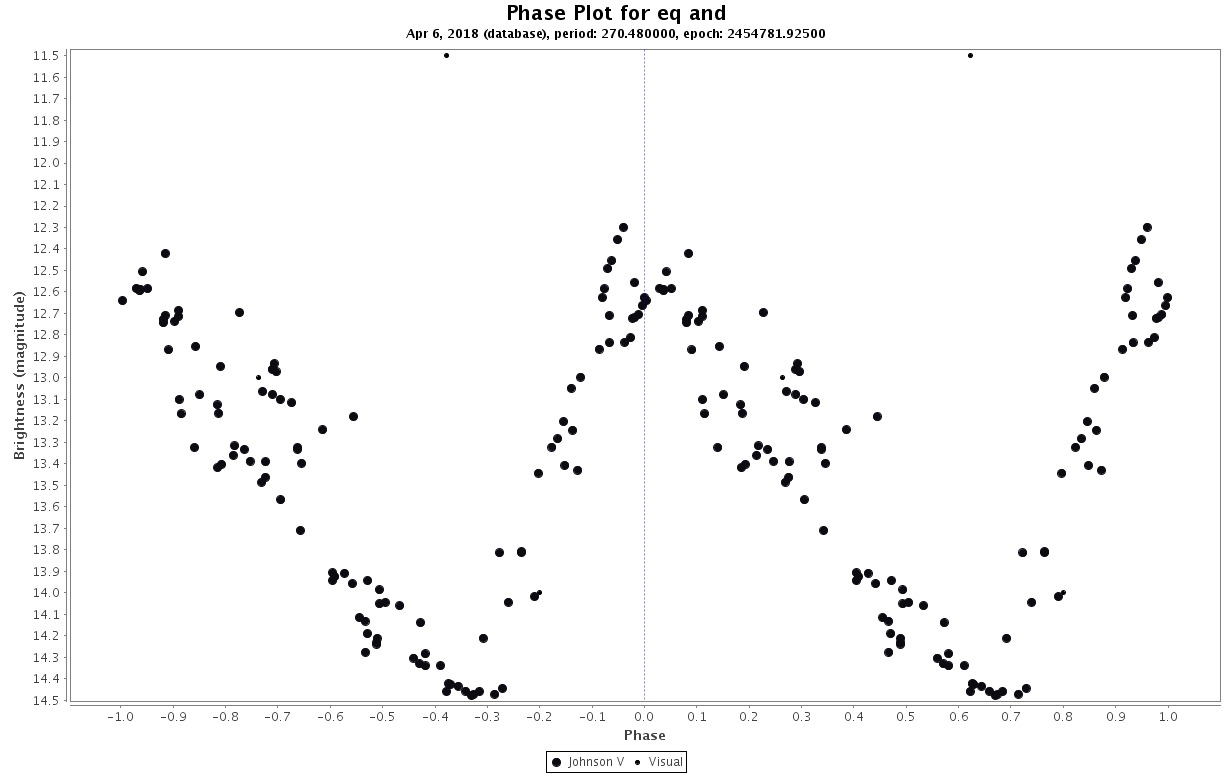}
\end{center}
\caption{EQ And is classified as SR. The phase diagram using visual
observations and a period of 270.48 days, shown here, is quite satisfactory.  The period
in VSX, 211: days, does not produce a good phase diagram.}
\end{figure}

\begin{figure}
\begin{center}
\includegraphics[height=7cm]{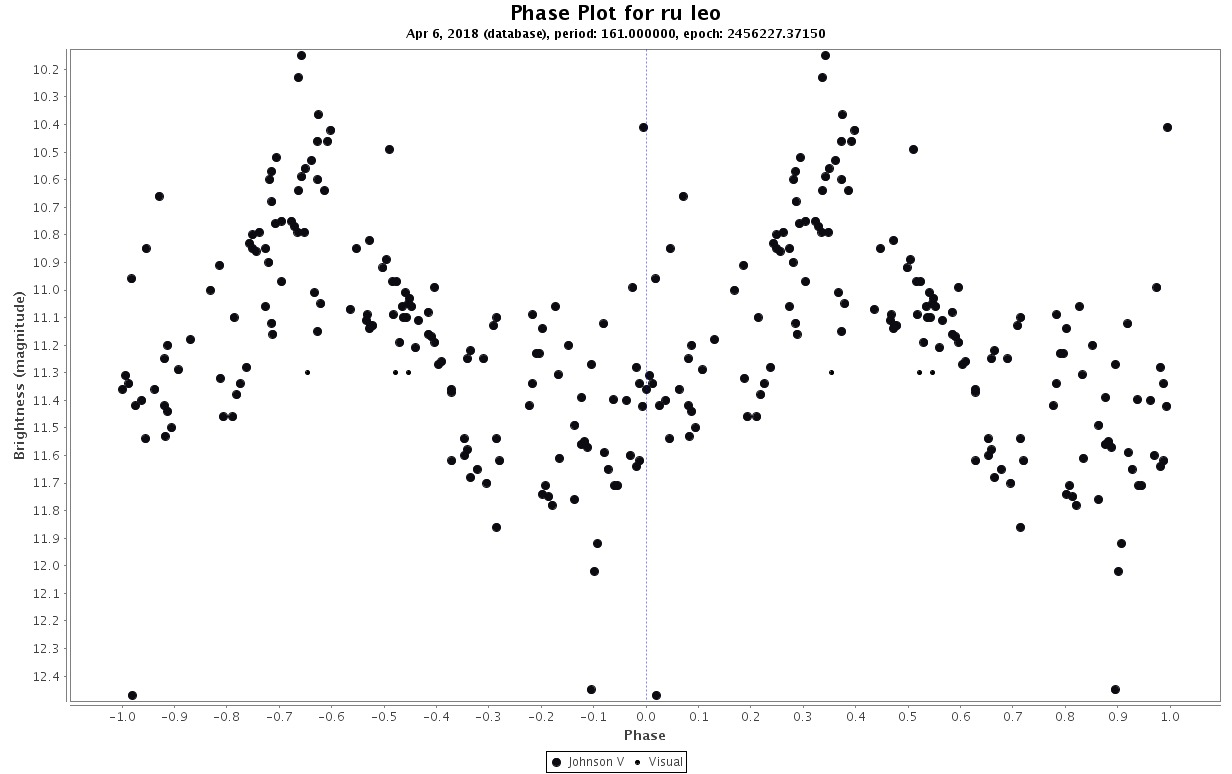}
\end{center}
\caption{RU Leo is classified as Lb (irregular).  The phase diagram using
Johnson V observations and a period of 161 days, shown here, is quite
satisfactory.}
\end{figure}

\medskip

\noindent
{\bf 3.1 Notes on Individual Stars}

\smallskip

These notes on individual SRa, SRb, SR, and Lb stars are combined,
and listed in order of constellation.  Many of the 156 stars in the
input list also have observations from other sources, such as ASAS, {\it
Hipparcos}, DIRBE (Smith {\it et al.} 2002), AFOEV etc. but, unless they helped in the analysis
or interpretation of the AAVSO data, they are not discussed here.

{\it KW Cep}: the Fourier spectrum is complex; the dominant cycle lengths
are about 150 days.

{\it UZ Cet:} the ASAS period is 80.9 days, but the average cycle length
is about 117 days.  PVSX, the new period, and the ASAS period produce
equally unsatisfactory phase curves, but PVSX is probably the best.

{\it RU CrB}: the V observations show cycle lengths of about 60 days,
but the early visual observations give a period of 436 days, which may
possibly be a long secondary period.

{\it VY Eri:} the ASAS light curve is very complex; irregular or
multiperiodic star?

{\it TZ Hor:} there is also a peak at 23.41 days, but the cycle lengths
are 35 days.

{\it DV Lac}: the Fourier spectrum is complex, with cycle lengths in
the range of 150 to 180 days.

{\it RS LMi:} complex; cycle lengths about 110 days; there may be a long
secondary period.

{\it V360 Peg}: the {\it General Catalogue of Variable Stars} (Samus
{\it et al.} 2017) classifies this star as possibly RV Tauri type,
but we find no evidence for this.

{\it SW Pic}: the V observations show periods of about 25 and 35 days.

{\it TW Ret}: the {\it General Catalogue of Variable Stars} (Samus {\it
et al.} 2017) classifies this star as possibly RV Tauri type, but we
find no evidence for this.

{\it BN Ser:} there are several peaks of comparable height in the
Fourier spectrum.  There appears to be a long secondary period.

\medskip

\noindent
{\bf 4. Discussion}

\smallskip

In the AAVSO observing program, there are 155 PRGs which are designated
as ``legacy stars", and recommended for regular observation.  Over the
last decade, they have averaged about 375 observations {\it per year}
(Pearce 2018).  These dense, sustained observations enable astronomers
to follow their wandering periods, variable amplitudes, multiperiodicity,
and long secondary periods (LSPs).

For the 156 stars in the present study, there are less than 250
observations {\it in total}.  As a result, less than half the stars
yield any meaningful result, that usually being only an estimate of the average
period and amplitude.  There are some stars which have only a few
dozen V observations obtained on a single night (!).  For others,
sparse observations are spread over many decades.  For others,
the Fourier spectra showed many comparable peaks, with none of them
prominent.  Some PRGs are known to pulsate in both the fundamental and
first-overtone modes.  There may be some stars for which the derived
period is actually an LSP, with the shorter pulsation period hidden in
the noise.  Some stars, especially the Lb stars, may be truly irregular.

A few of the Lb stars in Table 4 had only one or two cycles of V
observations, so it was not possible to say whether they showed any strict
periodicity.  A few others had more V observations, but not enough to say
whether they were multiperiodic or irregular.  A few, such as OR Cep,
RU Leo (Figure 2), and Z LMi, showed good phase curves, and may be SR.
Some of the stars in Tables 1-4 may, of course, have been misclassified as to
variable star type because of limited observations.

\medskip

\noindent
{\bf 5. Conclusions}

\smallskip

Of the 156 stars that were examined, less than half yielded any useful
information, that being an average period and amplitude.  In many cases,
that information was uncertain.  It is noteworthy, however, that about
a third of the Lb (irregular) variables showed some periodicity.

It is not clear that continued {\it sparse} observation of the 
stars in this study will yield better results.  And there are many more PRGs in the
AAVSO observing program with {\it less} than 150 observations in total,
and which were therefore not included in the present study.  AAVSO might wish to 
think seriously about how to manage these PRGs in its observing program.
If it is decided that these stars should continue to be observed,
then it might be best if observers ``adopted" stars for a year or two
(or three), to ensure that they were observed sufficiently regularly.

\medskip
\noindent
{\bf 6. Acknowledgements}

\smallskip

I thank the AAVSO observers who made the observations on which this project is based, the AAVSO staff who archived them and made them publicly available,
and the developers of the VSTAR package which was used in the analysis.
Special thanks to Elizabeth Waagen.  This project made use of the SIMBAD
database, maintained in Strasbourg, France.  The Dunlap Institute is funded
through an endowment established by the David Dunlap family, and the University of Toronto.

\medskip
\noindent
{\bf References}

\smallskip

\noindent
Benn, D. 2013, VSTAR data analysis software (http://www.aavso.org/vstar-overview)

\noindent
Kafka, S. 2018, variable star observations from the AAVSO International
Database:

 (https://www.aavso.org/aavso-international-database

\noindent
Pearce, A. 2018, {\it AAVSO Newsletter}, \#75, (January), 15.

\noindent
Percy, J.R. and Long, J. 2010, {\it JAAVSO}, {\bf 38}, 161.

\noindent
Percy, J.R. and Terziev, E. 2011, {\it JAAVSO}, {\bf 39}, 1.

\noindent
Percy, J.R. and Tan, P.J. 2013, {\it JAAVSO}, {\bf 41}, 1.

\noindent
Percy, J.R. and Kojar, T. 2013, {\it JAAVSO}, {\bf 41}, 15.

\noindent
Pojmanski, G. 1997, {\it Acta Astronomica}, {\bf 47}, 467.

\noindent
Samus, N.N. {\it et al.} 2017, {\it General Catalogue of Variable Stars}, Sternberg
Astronomical Institute, Moscow (GCVS database: http://www.sai.msu.ru/gcvs/gcvs/index.htm)

\noindent
Smith, B.J. {\it et al.} 2002, {\it Astron. J.}. {\bf 123}, 948.

\noindent
Templeton, M.R., Mattei, J.A., and Willson, L.A. 2005, {\it Astron. J}., {\bf 130}, 776.

\medskip

\begin{table}
\caption{Variability Properties of Some SRa Stars}
\begin{tabular}{rcrcl}
\hline
Star & PVSX (days) & P(days) & Amp (mag) & Notes \\
\hline

UV Aql & 385.5 & 350$\pm$30 & 0.17v & many peaks \\
LQ Ara & 183.7 & 179.73 & 0.34v & 1: \\
FX Cas & 289 & 292.4 & 1.44v & data consistent with PVSX \\ 
V533 Cas & 305 & 303.67 & 1.04v & 1 \\
V864 Cas & -- & 344 & 1.00V & P(vis) = 368 days \\
AL Cen & 125 & 128.7 & 0.69v & 1, ASAS P = 126.64 days \\
V343 Cep & 525 & 482.9 & 0.93v & 1 \\
UZ Cet & 121.74 & 203.34 & 0.14V & *; 1, multiperiodic? \\
V577 Cyg & 479 & 478.5 & 0.32v & 1, P(V) = 460.8 days \\ 
V659 Cyg & 514 & 509.68 & 0.73v & 1 \\
V1059 Cyg & 372 & 380$\pm$10 & 0.18v & poor phase curve; period spurious? \\ 
AY Her & 129.75 & 127.58 & 1.05v & 2 \\
IV Peg & 214.0 & 213.8 & 0.95v & 1, ASAS P = 210.387 days \\
TW Ret & 217.6 & 225.99 & -- & *, 2, RVT according to SIMBAD \\
VV Tel & 138.8 & 137.6 & 0.70V & 1 \\
UZ Vel & 354 & 390.6 & 0.16v & 1:, ASAS P = 353 days \\
AAVSO 0705+29 & 106.2 & 106.64 & 0.43v & 3 \\
\hline
\end{tabular}
\end{table}

\begin{table}
\caption{Variability Properties of Some SRb Stars}
\begin{tabular}{rcrcl}
\hline
Star & PVSX (days) & P(days) & Amp (mag) & Notes \\
\hline
W Ara & 122 & 119.6 & 0.14v & 1, ASAS P = 121.8 days \\
V505 Car & 26.5 & 20.266 & 0.02V & and/or 26.408 days \\
V481 Cas & 158.4 & 159 & 0.07V & \\
R Cir & 222 & 366.3? & 0.24v & 1, 3, ASAS P = 220 days \\
RU Crt & 60.85 & 700: & 0.39v & broad peak in Fourier spectrum \\
AQ Del & 71.9 & 71.6 & 0.16v & 1, ASAS P = 73.61 days \\
VY Eri & 102.5 & 189: & 0.23v & *; 1, ASAS P = 191 days, one cycle in V \\
V521 Ori & 221 & 225.17 & 0.36V & 1 \\
X Pav & 199.19 & 400.3 & 0.33v & 1 \\
V443 Per & 69.5 & 69.9 & 0.17V & 3, LSP $\sim$ 400 days \\
RW Psc & 154 & 154.2 & 0.15v & 2 \\
Z Ser & 88.2 & 88.3 & 0.18v & P(ASAS) = 89.379 days \\
BN Ser & 140.7 & same & 0.17V & *; ASAS P = 144.131 days \\
GK Vel & 120: & 182: & 0.08v? & 1, several peaks including 123.7 days \\
\hline
\end{tabular}
\end{table}

\begin{table}
\caption{Variability Properties of Some SR Stars}
\begin{tabular}{rcrcl}
\hline
Star & PVSX (days) & P(days) & Amp (mag) & Notes \\
\hline
EQ And & 211: & 273.6 & 0.82v & 1; see Figure 1 \\
KQ Aql & 164.2 & 417 & 0.50v & 2, PVSX gives good phase curve \\
V925 Aql & -- & 398.8 & 0.18v: & poor phase curve \\
SZ Ara & 221.8 & 219.8 & 0.77v & 1 \\
UW Cam & 544 & 523.3 & 0.33v & variable amplitude \\ 
AM Car & 314 & 408 & 0.45v & 1; also 50-day cycles \\
RU CrB & 436 & 427 & 0.2V & *; 2, 436 days may be an LSP \\
V1673 Cyg & 115.5 & 116.5 & 0.15v & \\
AE Del & 260 & 152.5 & 0.67v & PVSX is an alias \\
% NZ Gem & -- & 350 & 0.03v & also short period? \\
V529 Her & 400 & 197.3 & 0.10v & 2 \\
TZ Hor & ??? & 35.52 & 0.02V & *; also 23.41 days \\ 
Y Mic & 364: & 180$\pm$2 & 0.18v & 4 \\
V360 Peg & 44.9 & 45.28 & 0.09V & *; 1, RV Tau evidence weak \\
% V500 Per & -- & 55.7 & 0.26v & \\
V Pic & 180 & 173.3 & 0.62v & 1 \\
SW Pic & -- & 25.2$\pm$0.1 & 0.026V & * \\
$\gamma$ Ret & 25 & 29.87 & 0.034V & \\
DR Tuc & -- & 23.59 & 0.028V & {\it Hipparcos} P = 23.87 days \\
% FL Vel & 840 & 583 & 0.36V & \\
o Vir & -- & 30.50 & 0.036V & \\
NSV11453 & 153 & 296.47 & 0.85v & 1 \\
OOO-BLG-605 & 78.09 & same & 0.21V & 2 \\
\hline
\end{tabular}
\end{table}

\begin{table}
\caption{Variability Properties of Some Lb Stars}
\begin{tabular}{rccl}
\hline
Star & P(days) & Amp (mag) & Notes \\
\hline
KR Cep & 50 & 0.13V & one cycle in V \\
KT Cep & 77 & 0.16V & two cycles in V \\
KW Cep & 170$\pm$10 & 0.15v & *; complex; cycles 150 days long \\
OR Cep & 348.5 & 0.97v & good phase curve \\
% ST CrB & 80: & 0.3V & irregular; PVSX = 76.945743 days \\
DV Lac & 170$\pm$10 & 0.27V & *; irregular; result uncertain \\
PY Lac & 95 & 0.19V & one cycle in V; also short-period variability? \\
RU Leo & 161 & 0.38V & good phase curve \\
VX Leo & 95.6 & 0.16V & good phase curve, but complex, multiperiodic? \\
CP Leo & 190: & 0.07V & poor phase curve; complex, multiperiodic? \\
GK Leo & 345$\pm$5 & 0.16V & \\
Z LMi  & 161: & 0.31V & good phase curve \\
RS LMi & 90: & 0.13V & *; poor phase curve; complex \\
CX Mon & 385 & 0.35v & fair phase curve \\
WW Psc & 25$\pm$ & 0.03V & \\
FL Ser & 390$\pm$2 & 0.16v & fair phase curve \\
TT UMa & 490$\pm$10 & 0.1v & fair phase curve \\
NSV 623 & 74 $\pm$2 & 0.25v & uncertain; $\Delta$V = 0.50 \\ 
\hline
\end{tabular}
\end{table}

% SRa table

% SRb table

% SR table

% Lb table

\end{document}